# Two dimensional frustrated magnetic state in superconducting $RuSr_2Eu_{1.5}Ce_{0.5}Cu_2O_{10}$ (Ru-1222)

S. Garcia[1,2], L. Ghivelder[1], and I. Felner[3]

[1] Instituto de Física, Universidade Federal do Rio de Janeiro, C.P. 68528, Rio de Janeiro, RJ 21941-972, Brazil

[2] Instituto de Física, Universidade de São Paulo, C.P. 66318, São Paulo, SP 05315-970, Brazil

[3] Racah Institute of Physics, The Hebrew University, Jerusalem, 91904, Israel.

Abstract

In this paper we investigate the magnetic state and the role of the crystalline structure in $RuSr_2Eu_{1.5}Ce_{0.5}Cu_2O_{10}$ (Ru-1222). Measurements were made in the isomorphic series $(Nb_{1-x}Ru_x)Sr_2Eu_{1.5}Ce_{0.5}Cu_2O_{10}$ [(Nb,Ru)-1222], with $0 \leq x \leq 1$. 3D XY fluctuations above the magnetic transition were not observed in Ru-1222, suggesting a weak inter-plane coupling between the $RuO_2$ layers. The compositional dependence of the magnetic susceptibility shows a rapid broadening with increasing Nb content, explained in terms of a cluster-glass state. The variation of several superconducting parameters as a function of Ru content is linear in the whole concentration range, with no jumps at the critical concentration for which percolation of long–range order is expected. 3D Arrhenius- and Vogel-Fulcher-type dependencies fail to describe the dynamic properties. Fitting of a generalized Vogel-Fulcher-type dependence, with $\ln(\tau/\tau_0) = A(T-T_0)^{-B}$, yield B = 2.0, in excellent agreement with Monte Carlo simulations for 2D systems. The value deduced for $T_0$ agrees well with the re-opening of hysteresis in the M(H) curves. The observed superconducting and magnetic features are explained in terms of a scenario of 2D magnetic islands at the $RuO_2$ layers, with no long range magnetic order.

## 1 – Introduction

The magnetic structure of superconducting ruthenium-copper oxides, RuSr$_2$$R$Cu$_2$O$_8$ (Ru-1212) and RuSr$_2$($R$,Ce)$_2$Cu$_2$O$_{10}$ (Ru-1222), where $R$ = Gd, Eu, is still an open and controversial topic of investigation. Some features of this complex magnetic system include competitive antiferromagnetic (AFM) and ferromagnetic (FM) interactions,[1,2] itinerant magnetism of the Ru sub-lattice,[3] spin-flop transitions,[4] deviations from the Curie-Weiss behavior well above the magnetic transition temperature,[5] different states of valence for the Ru ions with possible ferrimagnetic order,[1] super-exchange interaction of the Dzyaloshinsky-Moriya type,[6] magnetic phase separation of nanosized particles,[7-9] and a dynamic response typical of a spin-glass state.[10,11] The details of the magnetic structure are of primary interest for the onset of the superconducting (SC) state at temperatures considerably below the magnetic transition. In addition, one must understand how a coherent SC order parameter can be established across the RuO$_2$ layers in spite of a exchange interaction energy between the spins and the conduction electrons with values of the order of the superconducting energy gap.[3]

Even in the relatively simpler case of the Ru-1212 system, different experimental techniques yield conflicting results about the nature of the magnetic order. Neutron powder diffraction (NPD) patterns[12] reveal a dominant G-type AFM order with a very small FM component (~ 0.1 $\mu_B$ per Ru ion), while a type-I AFM structure is more adequate to explain the observed magnetic properties.[1] On the other hand, nuclear magnetic resonance (NMR) spectra indicates the existence of a large FM component.[13] The magnetic behavior of Ru-1222 compounds is considerably more complex, and NPD measurements[14] have not yet provided a definitive answer about the nature of the magnetic order of the Ru moments. Recent reports provide more conflicting results: a long-range 3D magnetic order of the Ru moments at 140 K, with spin canting below 91 K, is claimed in RuSr$_2$Y$_{1.5}$Ce$_{0.5}$Cu$_2$O$_{10-\delta}$ [15]; but on the other hand a detailed diffraction study[16] yield no evidence of long-range magnetic order intrinsically associated to the Ru-1222 material, with complex patterns ascribed to unidentified impurity phases.

The aim of the present study is to address a basic question: is there an intrinsic long range magnetic order in Ru-1222? We carried out this investigation by studying the isomorphic $(Nb_{1-x}Ru_x)$-1222 series of compounds with $0 \leq x \leq 1$. Previous studies[17,18] in a limited concentration range ($x \geq 0.5$) showed that, as expected, the SC and magnetic transition temperatures decrease with the rise in Nb content. Nb-doping is particularly suitable to conduct a systematic study of how the magnetic properties emerge in Ru-1222, since Nb ions carry no magnetic moment and are pentavalent, near the +4.74(5) valence value for Ru ions in Ru-1222 (Ref. 19). The hole density on the SC planes decreases very slightly with increasing Nb-doping level and no change in the oxygen content was observed.[18] Another advantage is that superconductivity is present over the whole concentration range. By increasing the Ru content from dilute doping levels, the evolution of the magnetic response and its correlation with the changes in the superconducting properties may be followed. Our results show that in the measured samples there is no critical Ru concentration for percolation of long-range magnetic order, and that the Ru moments are coupled together in a frustrated two dimensional (2D) state.

## 2 – Experimental Details

Samples of $(Nb_{1-x}Ru_x)Sr_2Eu_{1.5}Ce_{0.5}Cu_2O_{10-\delta}$ with $x$ = 0, 0.2, 0.4, 0.5, 0.6, 0.7, 0.85, and 1.0 were prepared following the standard solid-state reaction technique. All samples have been prepared simultaneously under the same conditions. No spurious lines were observed in the x-ray diffraction patterns. Resistivity, dc magnetization and ac susceptibility measurements were performed in a Quantum Design PPMS system. Low frequency ac susceptibility, down to $f$ = 0.002 Hz, and low field magnetization, was measured with a Cryogenic SQUID magnetometer. Special care was devoted to measure "fresh" samples, immediately after oxygenation at 600 $^o$C for five days. A non superconducting sample with a different Ce content, $RuSr_2EuCeCu_2O_{10}$, was also investigated.

# 3 – Results and Discussion

Information about the dimensionality of the magnetic order and the orientation of the Ru moments in superconducting $RuSr_2Eu_{1.5}Ce_{0.5}Cu_2O_{10-\delta}$ and the isomorphic non superconducting $RuSr_2EuCeCu_2O_{10}$ was obtained by examining the fluctuations above the magnetic transition temperature, $T_M$. The value of $T_M$ was taken as the inflection point in the field cooled magnetization curve, as shown in Fig. 1. A log-log plot of the derivative of the zero-field-cooled dc susceptibility, $d\chi/dT$, plotted as a function of $[(T/T_M) – 1]$, , is shown in Fig. 2A linear dependence with a critical exponent $\gamma = 1.30$, corresponding to 3D XY fluctuations, as observed for Ru-1212 (Ref. 20), was not obtained. The smooth bump visible in the data corresponds to the contribution from a minority fraction of $Ru^{4+}$ ions (~10-15%),[2,21] due to slight deviations of oxygen stoichiometry with a local character. This component is superimposed to the general background coming from the majority fraction of $Ru^{5+}$ ions. The isomorphic non-superconducting $RuSr_2EuCeCu_2O_{10}$ compound, where all the Ru ions are pentavalent ,[21] was measured in order to eliminate the $Ru^{4+}$ contribution. These results are also shown in Fig. 2. The non linear dependence obtained indicates that the absence of 3D XY fluctuations is a characteristic of the $RuO_2$ layers in the Ru-1222 structure. These results point to weak or zero interplane coupling and a considerable lower out-of-plane anisotropy in comparison to Ru-1212. As discussed below, we obtained further evidence supporting the decoupling between the $RuO_2$ layers, therefore it is very unlikely that a different 3D critical behavior would apply for Ru-1222. The relevant structural difference between Ru-1212 and Ru-1222 is the large separation between the $RuO_2$ layers and the misalignment of the superexchange chain in the former [22], strongly affecting the coupling; the Ru-O bond distances and the tilting angles of the $RuO_6$ octahedra are very close in both systems. This strongly suggests that the origin of the difference in the fluctuation behavior is associated to an enhanced 2D character of the crystalline structure in Ru-1222.

In Fig 3, selected resistive curves in the region of the SC transition are presented, normalized to the value at T = 70 K. The corresponding derivative curves, shown in the inset, exhibit two peaks. As Nb substitutes Ru in Ru-1222, we may follow the changes in the SC transition temperature, $T_{SC}$, taken as the value for which $d\rho/dT = 0$, and of the

intragrain transition temperature, $T_{intra}$, as determined from the peaks at the higher temperature. The peaks at lower temperatures correspond to the intergranular transition.[4] The compositional dependence of $T_{SC}$ and $T_{intra}$ is presented in Fig. 4.

Figure 5a shows ac susceptibility curves, $\chi'(T)$, measured for the $(Nb_{1-x}Ru_x)$-1222, with $x = 0.6, 0.7, 0.85$, and $1.0$. Zero field cooled low field dc magnetization measurements (not shown) yield very similar results. As the Ru content $x$ decreases, the curves rapidly broaden. For $x = 0.85$, the peak of the magnetic transition can be still identified at around $T = 60$ K. For this composition the contribution to the ac susceptibility coming from the Ru-sub-lattice is dominant, and the emergence of the SC state on cooling is revealed by shoulders and changes in slope for $T \leq 40$ K. For $x = 0.7$ and $0.6$, the relative strength of the magnetic and SC components is reverted. The magnetic transition is so smeared that it can no longer be detected as a peak. Instead, it appears as a broad contribution above the maximum at $T \approx 28$ K, which corresponds to the intergranular SC transition peak (see the inset in Fig. 3 for $x = 0.6$). For comparison, in $(Sn_{1-x}Ru_x)$-1212 [Ref. 23] and $(Nb_{1-x}Ru_x)$-1212 [Ref. 24], the magnetization curves are shifted as the Ru-sub-lattice is diluted, allowing the determination of the compositional dependence of $T_M$, which is not possible for the curves in Fig. 5(a). However, their derivatives, presented in Fig. 5(b), give relevant information about how the coupling between the Ru moments evolves. For Ru-1222 ($x = 1.0$), a well defined minimum clearly marks $T_M = 85$ K. For $x = 0.85$, a broad minimum at $T = 73$ K is observed, with a shoulder at $T_M$ ($x = 1.0$). As $x$ decreases further, although the magnetic transition is smeared over several tens of degrees, the shoulder is still present. The shallow minima around $T = 38$ K for the four compositions considered in Fig. 5(b), agree well with their corresponding intragrain SC peaks in the inset of Fig. 3, and are not linked to the magnetic transition. The maxima for $T \leq 22$ K correspond to the zero resistance temperature for the SC transition (see Fig. 3).

The features of the $d\chi'/dT$ curves can be understood in a scenario of magnetic frustration enhanced by Nb doping. We have shown before[2] that magnetic frustration is an intrinsic property of Ru-1222 ($x = 1.0$), described by a cluster-glass state with short-range correlations of the Ru moments inside a coherence volume of nanometric size. As the spin system is diluted by Nb doping, the superexchange coupling between the Ru

moments surrounding the Nb sites is weakened, enhancing magnetic frustration. The spatial distribution of the Nb ions generates an array of regions or islands inside which the magnetic state is identical to the parent Ru-1222 compound. The contribution from these islands to the susceptibility is the origin of the shoulder observed at $T_M$. On cooling, neighboring islands become gradually coupled, leading to the broadening of the magnetic transition. For higher Nb concentrations, the exact topology of the islands is more complex than a simple picture of perfect regions separated by a network of Nb ions, but the basic idea remains: a gradual nucleation of the cluster-glass state across the sample as the temperature diminishes, accompanied by a temperature independent shoulder. Magnetic frustration is the key point to understand the phenomena. Therefore, the magnetic response of the diluted Ru-spin system in Ru-1212,[23,24] exhibiting long-range magnetic order, is not at all similar to the results of the present study.

The compositional dependence of the SC parameters for the $(Nb_{1-x}Ru_x)$-1222 series supports the absence of long-range magnetic order in Ru-1222. As the Ru content gradually increases from $x = 0$, one could expected that above some critical Ru concentration a jump or discontinuity in both $T_{SC}$ and $T_{intra}$ would signal the onset of long-range magnetic order in the compound. Instead, we observe a linear increase of both critical temperatures over the whole composition interval, as shown in Fig. 4. In addition, the difference $\Delta T = T_{SC} - T_{intra}$, smoothly increases from 2.2 K for $x = 0$ (Nb-1222) up to 4.7 K for $x = 1.0$ (Ru-1222), as observed in Fig. 4. This parameter is related to strong intragrain granularity associated to the effects of magnetic ordering on the onset of a coherent SC order parameter across the grain.[4,25] Therefore, the intragrain granularity effects are also gradually enhanced with the increase in $x$. Other SC parameters, such as the value of the applied magnetic field at the minimum in the virgin M(H) curves, $H_{min}$, smoothly decreases with the rise in $x$ (Fig.4, inset). The scenario of a frustrated cluster-state inside regions separated by Nb ions is consistent with the observed continuous change in the superconducting parameters, since the rise in the Ru content will simply gradually increase the size of the regions and their coupling on cooling.

The results presented above point to weak or zero interplane interaction between the $RuO_2$ layers. It has been proposed that these layers are only weakly coupled through dipole-dipole interactions,[26] suggesting a magnetic structure with an enhanced 2D-

character. In a previous report[2], we studied the scaling behavior of the magnetization and the shift in temperature of the $\chi'$-peaks in ac susceptibility measurements as the frequency of the driving field is varied. The results obtained are used in the present study as a basis to determine the dimensionality of the magnetic state. The scaling behavior evidenced that below $T_M$ a magnetic cluster-glass state emerges, with coupling between the clusters on cooling. In relation to the dynamic response, the shift in the $\chi'$-peaks can not be described by a thermally activated process of the Arrhenius- and Vogel-Fulcher-type process. The inadequacy of the Vogel-Fulcher dependence was revealed by the deviation from linearity of the $-1/ln[2\pi f/f_0]$ versus $T_p$ plot, where $T_p$ marks the temperature for the maximum in the $\chi'$-curves; the experimental data collected in Ref. 2 are reproduced in Fig. 6.

In an attempt to determine if the dynamic response has a 2D-like behavior, we considered a generalized Arrhenius dependence used to characterize thermal activation processes in 2D-spin-glass systems,[27]

$$\ln(\tau/\tau_0) = A\, T^{-(1+\psi\nu)}, \qquad (1)$$

where $\psi$ and $\nu$ are the barrier and correlation-length exponents, respectively, with $\tau = 1/f$ and $\tau_0 = 1/f_0$. Such systems have been successfully described in terms of expression (1), with $1 + \psi\nu$ ranging from 2 (Ref. 28) to 2.5 (Ref. 29, 30). This generalized Arrhenius dependence is derived from the droplet scaling theory of Fisher and Huse,[31] and Monte Carlo simulations in 2D systems have given values of $(1 + \psi\nu) \sim 2$ (Ref. 32, 33). However, we failed to obtain a good fit of expression (1) to our $f$ vs. $1/T_p$ data, keeping $(1 + \psi\nu)$ in the 2 – 2.5 interval. It must be noted that the failure of the generalized Arrhenius model does not imply that the magnetic structure is not 2D in nature, in the same way that the inadequacy of the conventional (3D) Arrhenius law in 3D magnetic systems with interacting clusters does not mean that they are not 3D.

Following the idea of Eq. (1), we considered a similar generalized Vogel-Fulcher type equation of the form

$$\ln(\tau/\tau_0) = A\,(T - T_0)^{-B}, \qquad (2)$$

which yield a dependence of the type

$$-1/\ln[2\pi f/f_0] = (1/A)(T-T_0)^B \qquad (3)$$

The fit of expression (3) to the data in Fig. 6 yield B = 2.0, with $T_0$ = 78 K. In principle, it is not obvious that the value of the exponent for the 2D generalization of the Vogel-Fulcher law should be the same as the one for Eq. (1). There is close relation between the conventional Arrhenius and Vogel-Fulcher laws in the characterization of the dynamic response of variety of physical phenomena, where frustrated long range order is accompanied by coupling between entities with short-range correlations. The continuous transition from the Arrhenius law to the Vogel-Fulcher one in spin glasses[34] is clearly revealed by the progressive change of the time dependence of the freezing temperature as the strength of the coupling rises by increasing the concentration of magnetic ions, in a similar fashion as for the $(Nb_{1-x}Ru_x)$-1222 system. The generality and universal character of the Vogel-Fulcher dependence as a natural extension of the Arrhenius behavior have been discussed in other studies.[35] The coupling between the interacting entities is properly accounted by the simple introduction of the phenomenological parameter $T_0$ in the Arrhenius expression, with no change in exponent in any case. These results suggest that the same would hold for the 2D case. Monte Carlo studies of the spin dynamics for 2D-spin glasses considering interacting clusters confirm this assumption. The temperature dependence of the relaxation time and the distribution of the average effective energy barrier for a system of frustrated plaquettes were found to be properly described by a Vogel-Fulcher-type dependence.[37] It was also demonstrated[38] that the ac susceptibility evaluated in terms of the relaxation time obtained using the Vogel-Fulcher dependence reproduces a reasonable frequency dependence of the freezing temperature determined by the cusp in the susceptibility. In these works, however, there is not an explicit expression for the rate at which the relaxation time should diverge on cooling [i.e., the value of the exponent in Eq. (2)]. A more developed model[39] based on scaling considerations of the auto-correlation spin function demonstrated that the relaxation time obtained from the scaling is found to diverge exponentially with exponent proportional to $T^{-2}$. These considerations support Eq. (2) as the valid 2D-generalization of the Vogel-Fulcher dependence.

We analyze now the value deduced for $T_0$. Since in the conventional Vogel-Fulcher dependence this parameter is related to the inter-particle interactions, we measured the M(H) loops for different temperatures, looking for a signature of the assumed coupling between the clusters in the coercive field $H_c$. The $H_c(T)$ dependence obtained is shown in the inset of Fig. 6. For $T \leq T_M$, the relevant feature is the re-opening of hysteresis at ~70 K, close to the value deduced for $T_0$, which supports the generalized 2D Vogel-Fulcher type fit. The $H_c(T)$ dependence above $T_M$ is determined by a minority fraction (~10 at. %) of nano-sized $Ru^{4+}$-rich islands, generated by slight local deviations of oxygen stoichiometry.[21] The absence of long-range magnetic order as the concentration of Ru is increased in $(Nb_{1-x}Ru_x)$-1222 does not agree with the results for diluted quasi-2D magnetic systems,[40] either of the Ising or Heisenberg spin type. In such systems, a critical concentration $x_C = 0.59$ of magnetic ions for percolation of long-range order has been observed, as predicted for a 2D square lattice.[41,42] We found no evidence of such percolation in the ac susceptibility curves presented in Fig. 4(b), where a continuous evolution for $x \geq 0.6$ is observed. The quasi-2D magnetic compounds undergoes 3D ordering due to slight deviations from an ideal 2D system, associated to interlayer coupling [40] or spin anisotropy of XY symmetry, [43] and that an ideal 2D-Heisenberg system does not develop long-range order at finite temperatures. Therefore, the large separation between the $RuO_2$ layers in the Ru-1222 structure and the absence of, or very small, XY anisotropy, as revealed by the lack of 3D XY fluctuations, makes this compound a real 2D magnet. Further work is needed to establish a consistent correlation between the observed 2D magnetic properties and neutron diffraction results, which are contradictory. Such discrepancies are found elsewhere: in other rare-earth-ruthenium oxides, such as the pyrochlore $R_2Ru_2O_7$ systems, neutron diffraction experiments reveal long-range antiferromagnetic order of the Ru moments, while the bulk magnetic properties show a spin-glass-like behavior.[44]

The magnetic structure of Ru-1222 is possibly related to the fact that $T_{sc}$ is about 10 K lower than for Ru-1212. It has been proposed that Ru-1212 is a natural system to form the so called π-phase SC order parameter, predicted for SC-FM superlattices, presenting a node at the $RuO_2$ layers with a strong decrease of pair-breaking effects.[45] The emergence of a π-phase state is only possible if the magnetization in the magnetic layers exceeds a certain critical value.[46] The magnetization at the $RuO_2$ planes in Ru-1212, exhibiting long-range magnetic order, is about 4kG, [45] favoring π-phase formation. The 2D frustrated magnetic state proposed for Ru-1222 would prevent the RuO$_2$ layers to effectively work as π-junctions. Taking into account that both Ru-1212 and Ru-1222 have essentially the same rotations of the RuO$_6$ octahedra and quite similar Ru-O-Ru and Ru-O-Cu bond lengths, [47] it is unlikely that the difference observed in T$_{SC}$ would be due to structurally induced changes in the CuO$_2$ conduction band.

## 4 – Conclusions

In summary, conclusive evidence was provided showing that the magnetic response of Ru-1222 is consistent with a scenario of 2D interacting clusters at the RuO$_2$ layers, with no long-range magnetic order. Diluting Ru-1222 with Nb-1222 allowed a detailed study of how the magnetic order changes as a function of the concentration of Ru ions. The correlation with the changes induced in the superconducting parameters, quantitative determinations of the parameters characterizing the dimensionality of the dynamic properties, and their consistency with the dc results, strongly support the scenario of a two dimensional frustrated magnetic state. The large separation between the RuO$_2$ layers is the key feature determining the magnetic response of the system..

## Acknowledgments

This research was funded by FAPERJ, CNPq, Israel Science Foundation (ISF, 2004 grant number: 618/04), and Klachky Foundation for Superconductivity. S.G. acknowledges financial support from CLAF and FAPESP (Grant No. 07/51458-6).

**Figure Captions**

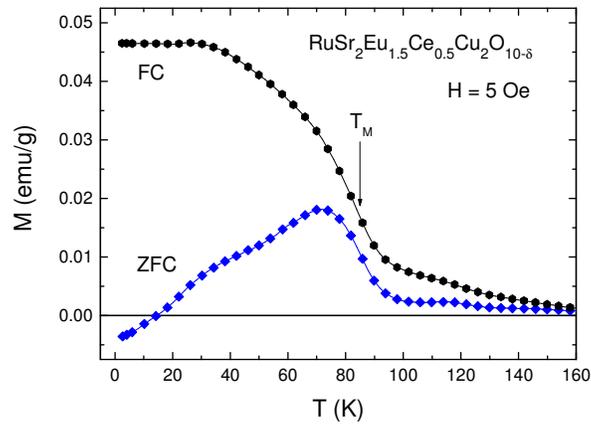

**Fig. 1** (Color online) Zero field cooled (ZFC) and field cooled (FC) dc magnetization of $RuSr_2Eu_{1.5}Ce_{0.5}Cu_2O_{10-\delta}$ (Ru-1222), measured with H = 5 Oe. The value of $T_M$ is indicated in the FC curve.

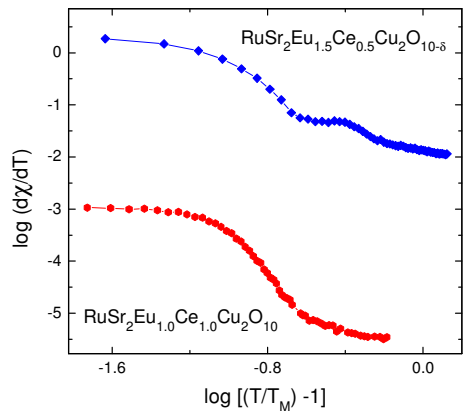

**Fig. 2** (Color online) Log-log plot of the derivative of the zero field cooled susceptibility, $d\chi/dT$, plotted as a function of $[(T/T_M) - 1]$ above the magnetic transition temperature, $T_M$, for $RuSr_2Eu_{1.5}Ce_{0.5}Cu_2O_{10-\delta}$ and the isomorphic compound $RuSr_2Eu_{1.0}Ce_{1.0}Cu_2O_{10-\delta}$.

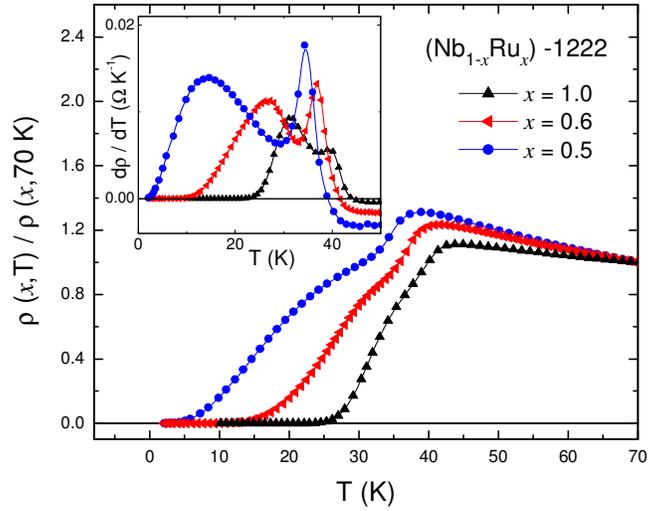

**Fig. 3** (Color online) Resistive superconducting transition for selected samples of $(Nb_{1-x}Ru_x)Sr_2Eu_{1.5}Ce_{0.5}Cu_2O_{10-\delta}$. The curves are normalized to the value at T = 70 K. Inset: the corresponding derivative curves, $d\rho/dT$.

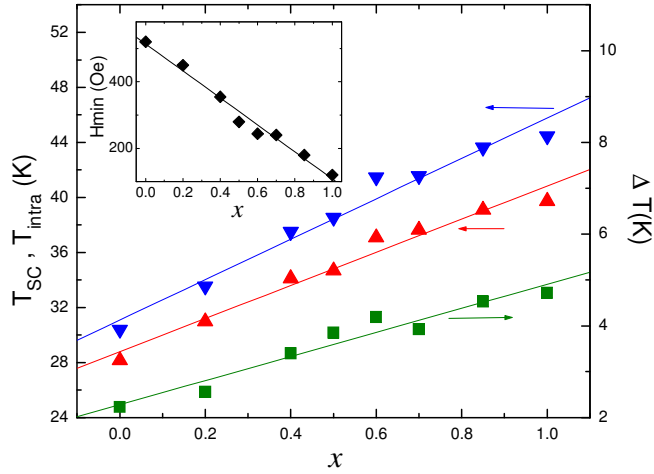

**Fig. 4** (Color online) The dependence of the onset temperature of the superconducting state, $T_{SC}$ (▼), the intragrain transition temperature, $T_{intra}$ (▲), and the difference $\Delta T = T_{SC} - T_{intra}$ (■), with the Ru content $x$ for the $(Nb_{1-x}Ru_x)Sr_2Eu_{1.5}Ce_{0.5}Cu_2O_{10-\delta}$ system. Inset: the compositional dependence of the applied magnetic field for the minimum (diamagnetic) net magnetization, $H_{min}$, in the virgin M(H) branches. The straight lines are linear fits.

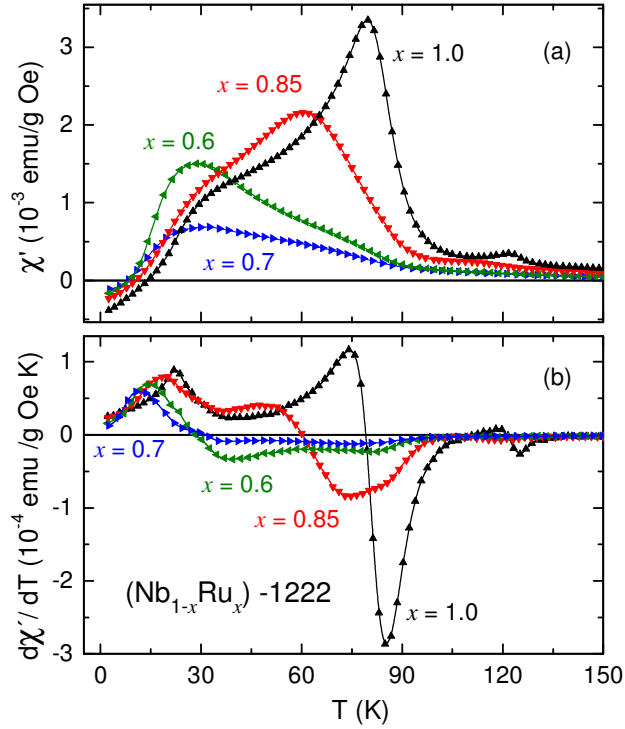

**Fig. 5** (Color online) (a) The ac susceptibility, $\chi'$, of $(Nb_{1-x}Ru_x)Sr_2Eu_{1.5}Ce_{0.5}Cu_2O_{10-\delta}$ with $x$ = 0.6, 0.7, 0.85, and 1.0 and (b) the corresponding derivatives, $d\chi'/dT$. Measured with an ac field of 5 Oe, and frequency of 1 kHz.

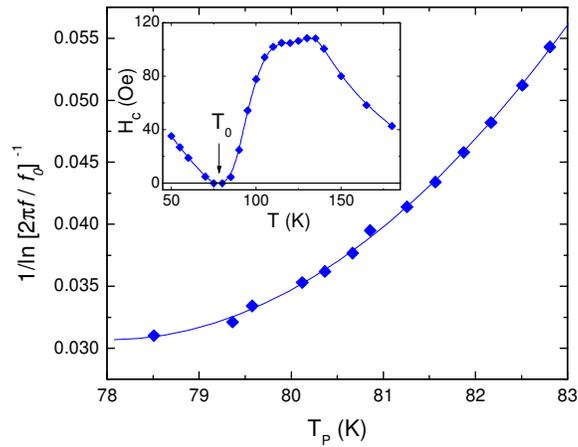

**Fig. 6** (Color online) The dependence of $1/\ln[(2\pi f/f0)]^{-1}$ with the $\chi'$-peak temperature $T_p$ for $RuSr_2Eu_{1.5}Ce_{0.5}Cu_2O_{10-\delta}$, using $f_0 = 10^{12}$ Hz; $f$ is the frequency of the driving field, which varies approximately from 0.002 to 3000 Hz. The continuous line is the best fit of Eq. (3). Inset: the temperature dependence of the coercive field $H_c$. The phenomenological parameter $T_0$ deduced from the fit is indicated (see text for details).